\documentclass[reprint,showpacs,preprintnumbers,amsmath,amssymb,prl,twocolumn]{revtex4}
\usepackage[pdftex]{graphicx}
\usepackage{bm}
\usepackage{float}
\usepackage{color}
\usepackage{hyperref}
\usepackage{pdfpages}
\usepackage[T1]{fontenc}
\begin{document}
\title{Characteristics of the secondary relaxation process in soft colloidal suspensions}
\author{Debasish Saha}
\email{debasish@rri.res.in}
\affiliation{Soft Condensed Matter Group, Raman Research Institute, C. V. Raman Avenue, Sadashivanagar, Bangalore 560 080, INDIA}
\author{Yogesh M Joshi}
\email{joshi@iitk.ac.in}
\affiliation{Department of Chemical Engineering, Indian Institute of Technology Kanpur, Kanpur 208 016, INDIA.}
\author{Ranjini Bandyopadhyay}
\email{ranjini@rri.res.in}
\affiliation{Soft Condensed Matter Group, Raman Research Institute, C. V. Raman Avenue, Sadashivanagar, Bangalore 560 080, INDIA}
\vspace{0.5cm}
\date{\today}
\begin{abstract}
A universal secondary relaxation process, known as the Johari-Goldstein (JG) $\beta$-relaxation process, appears in glass formers. It involves all parts of the molecule and is particularly important in glassy systems because of its very close relationship with the $\alpha$-relaxation process. However, the absence of a J-G $\beta$-relaxation mode in colloidal glasses raises questions regarding its universality. In the present work, we study the microscopic relaxation processes in Laponite suspensions, a model soft glassy material, by dynamic light scattering (DLS) experiments. $\alpha$ and $\beta$-relaxation timescales are estimated from the autocorrelation functions obtained by DLS measurements for Laponite suspensions with different concentrations, salt concentrations and temperatures. Our experimental results suggest that the $\beta$-relaxation process in Laponite suspensions involves all parts of the constituent Laponite particle. The ergodicity breaking time is also seen to be correlated with the characteristic time of the $\beta$-relaxation process for all Laponite concentrations, salt concentrations and temperatures. The width of the primary relaxation process is observed to be correlated with the secondary relaxation time. The secondary relaxation time is also very sensitive to the concentration of Laponite. We measure primitive relaxation timescales from the $\alpha$-relaxation time and the stretching exponent ($\beta$) by applying the coupling model for highly correlated systems. The order of magnitude of the primitive relaxation time is very close to the secondary relaxation time. These observations indicate the presence of a J-G $\beta$-relaxation mode for soft colloidal suspensions of Laponite.
\end{abstract}
\maketitle
\section{Introduction}
The exact nature of the relaxation processes in complex, disordered and out-of-equilibrium systems is one of the many unsolved problems in non-equilibrium statistical mechanics. The plethora of non-equilibrium systems around us supply several opportunities to observe diverse fascinating phenomena. The physics of relaxation processes in glasses or glassy systems has been meticulously studied during the last century. It has been noted that glasses comprise a disordered state of matter which are structurally like liquids, but whose dynamics are characterized by extremely slow relaxation. These systems fail to relax within timescales accessible in the laboratory. At the particulate scale, many types of relaxation processes are possible that can involve an atom, a molecule, a part of molecule, a group of molecules or particles. It is important to study how these microscopic relaxation processes slow down as glass-forming materials are quenched below their glass transition temperatures, with their relaxation time scales exceeding the observation time scales as the system falls out of equilibrium.\\
\indent The different types of relaxation processes in glass formers can be classified in two categories- the primary and the secondary relaxation processes \cite{Gotze_Sjogren_RPP_1992,Debenedetti_Stillinger_Nature_2001}. The former is involved with the orientational or structural rearrangements of the molecules or particles and is considered to be the main process leading to structural relaxation. This type of structural relaxation process, ubiquitous in molecular glasses or supercooled liquids, metallic glasses, polymer glasses and other glass formers, is generally known as the $\alpha$-relaxation process \cite{Gotze_Sjogren_RPP_1992}. Secondary relaxation processes, also known as $\beta$-relaxation processes, are much faster than the $\alpha$-relaxation process and involve the motion of a molecule or a part of a molecule and are generally believed to have no connection with the glass transition process \cite{Ngai_Paluch_JCP_2004}. However, many experimental results have shown that some $\beta$-relaxation processes are closely related to the structural relaxation process. Very recent molecular dynamics simulation results for asymmetric diatomic molecular glass formers demonstrate that the $\alpha$-relaxation process has a close relationship with a secondary relaxation process known as the Johari-Goldstein $\beta$-relaxation (J-G) process \cite{Fragiadakis_Roland_PRE_rapid_comm_2012,Fragiadakis_Roland_PRE_2013,Fragiadakis_Roland_PRE_2014,Fragiadakis_Roland_PRE_2015}. This type of secondary relaxation process is generally considered to be universal in nature as it appears in a variety of glass formers such as supercooled liquids, metallic glasses, polymeric glasses and plastic crystals. It involves all parts of a molecule or a particle and is particularly important in glassy systems because of its very close relationship with the $\alpha$-relaxation process. However, the absence of any experimentally reported result involving the detection of J-G $\beta$-relaxation mode in colloidal glasses has raised questions regarding its universality.\\
\indent After decades of research, a general route is used to classify the secondary relaxation processes in glass formers based on their dynamic properties \cite{Ngai_Paluch_JCP_2004} and their relation to the structural relaxation process. These include a thorough characterization of the glass former in terms of the following criteria: (i) whether the molecule possesses internal degrees of freedom; (ii) the involvement of the whole molecule or part of a molecule in the secondary relaxation process; (iii) the separation between $\tau_{\beta}$ and $\tau_{\alpha}$, the primary or structural relaxation times; (iv) the relationship between $\tau_{\beta}$ and $\tau_{0}$, where $\tau_{0}$ is the independent primitive relaxation time according to the coupling model; (v) the temperature dependence of the secondary relaxation strength $\Delta\epsilon_{\beta}$ (measured from dielectric spectroscopy) across the glass transition temperature $T_{g}$; (vi) the aging behavior of $\tau_{\beta}$ and $\Delta\epsilon_{\beta}$ below $T_{g}$; (vii) merging of $\tau_{\beta}$ with the primary relaxation time $\tau_{\alpha}$ at very high temperatures.\\ 
%
\indent We study the microscopic relaxation processes for Laponite suspensions, a model soft glassy material, by dynamic light scattering (DLS) experiments. A Laponite particle is disk-shaped (diameter $=25-30$ nm and thickness $\approx 1$ nm) and belongs to the 2:1 phyllosilicates family with layered hydrous magnesium silicate. Aqueous suspension of Laponite is proposed to form different phases, e.g. Wigner glass, gel, nematic, attractive glass etc., depending on clay concentration, ionic strength and waiting time after preparation \cite{Ruzicka_Soft_Matter_2011}. In this work, the $\alpha$ and $\beta$-relaxation timescales are estimated from the autocorrelation functions obtained in DLS measurements for Laponite suspensions with different concentrations ($C_{L}$), salt concentrations ($C_{S}$) and temperatures ($T$). We measure the primitive relaxation timescales from the $\alpha$-relaxation time and the stretching exponent $\beta$ by applying the coupling model for highly correlated systems. Our experimental results suggest that the $\beta$-relaxation process involves all the parts of the Laponite particle and is coupled with the primitive relaxation process. The ergodicity breaking time is shown to be correlated with the activation energy of the $\beta$-relaxation process for all $C_{L}$, $C_{S}$ and $T$. Our studies indicate that the $\beta$-relaxation process of colloidal glasses of Laponite demonstrates several similarities with the J-G $\beta$-relaxation processes and may, indeed, be a J-G $\beta$-relaxation mode.       
\section{Sample preparation and Experimental methods}
Laponite (Laponite RD $\circledR$) is procured from BYK Additives, Inc. Laponite powder is dried for up to 16 hours at $120^{\circ}$C in an oven to remove absorbed water. Dried Laponite is added to Millipore water (resistivity 18.2 M$\Omega$-cm) and is stirred vigorously using a magnetic stirrer until the suspension appears optically clear. The suspension is filtered at a constant flow rate of 3.0 ml/min using a syringe pump (Fusion 400, Chemyx Inc.) and a 0.45 $\mu$m Millipore Millex-HV syringe-driven filter unit. A sodium chloride (NaCl procured from Sigma-Aldrich) solution of a predetermined concentration is prepared and is added using a pipette to the filtered Laponite suspensions. Each suspension is vigorously stirred for 5-10 minutes to mix the salt solution. The sample is filled in a cuvette and then sealed with wax and tape for the DLS experiments. The waiting time or the aging time, $t_{w}$, is calculated from the moment the sample is sealed. A Brookhaven Instruments Corporation (BIC) BI-200SM spectrometer is used to study the samples. A 532 nm solid state laser (NdYVO$_{4}$, Coherent Inc., Spectra Physics) of output intensity 150 mW is used in the experiments to achieve a high scattered photon count. The scattering angle $\theta$ is fixed at $90^{\circ}$ for all time evolution experiments reported here. The temperature $T$ of the suspension is controlled by water circulation with a temperature controller (Polyscience Digital) attached to the DLS setup. The intensity autocorrelation function of the scattered light, $g^{(2)}(t)$, defined as $g^{(2)}(t)=\frac{<I(0)I(t)>}{<I(0)>^{2}}=1+A|g^{(1)}(t)|^{2}$, is measured using a digital autocorrelator (Brookhaven BI-9000AT) \cite{bern_pecora}. Here, $I(t)$ is the intensity of the scattered light at a delay time $t$, $g^{(1)}(t)$ is the normalized electric field autocorrelation function and $A$ is the coherence factor. The concentration of Laponite, expressed in \% w/v, refers to the weight of Laponite powder in grams that is mixed in 100 ml (100 gm) of water.
\section{Results and discussions}
\begin{figure*}[!t]
\begin{center}
\includegraphics[width=6.9in]{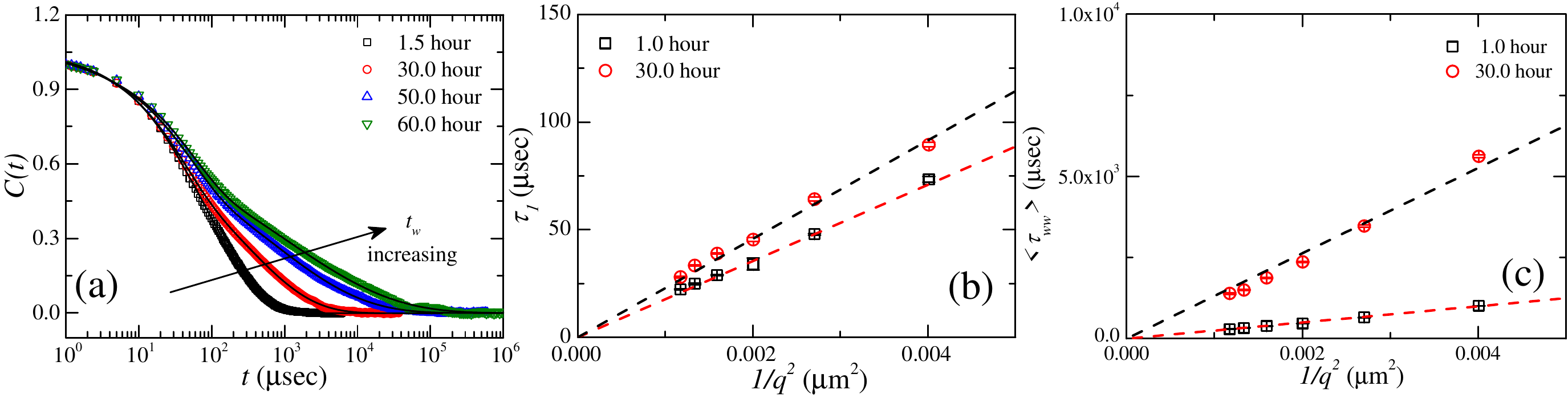}
\caption{(a) The normalized intensity autocorrelation functions $C(t)$ {\it vs.} the delay time $t$ at 25$^{\circ}$C and at a scattering angle $\theta$=90$^{\circ}$ for a 2.5\% w/v Laponite suspension at several different waiting times $t_{w}$ (from left to right): 1.5 hour ($\square$), 30.0 hour ($\circ$), 50.0 hour ($\triangle$) and 60.0 hour ($\nabla$). The solid lines are fits to equation~\ref{autocorrelation_function_secondary}. The diffusive dynamics of the fast relaxation time ($\tau_{1}$) and the mean slow relaxation time ($<\tau_{ww}>$) are shown in (b) and (c) respectively for a 2.5\% w/v Laponite sample for two different waiting times $t_{w}$. The dashed lines are linear fits passing through the origin.}
\label{autocorrelation_secondary}
\end{center}
\end{figure*}

Intensity autocorrelation functions $g^{(2)}(t)$ are acquired in dynamic light scattering experiments (DLS) as a function of the waiting time $t_{w}$ of aging Laponite suspensions. In figure~\ref{autocorrelation_secondary}(a), we plot the normalized intensity autocorrelation function, $C(t) = g^{(2)}(t)-1$, as function of delay time $t$ for several waiting times for a 2.5\% w/v Laponite suspension at 25$^{\circ}$C and at a scattering angle $\theta=90^{\circ}$. It is seen from this figure that the $C(t)$ data show two-step relaxations and that the slow relaxation processes progressively slow down with $t_{w}$. The two-step relaxation process in $C(t)$ can be expressed as the sum of an exponential and a stretched exponential decay \cite{Ruzicka_JPCM_2004,Ruzicka_PRL_2004,Debasish_YMJ_Ranjini_Soft_Matter_2014}.\\
\begin{equation}
\label{autocorrelation_function_secondary}
C(t)=[a\exp{\left\{-t/\tau_{1}\right\}}+(1-a)\exp{\left\{-(t/\tau_{ww})^{\beta}\right\}}]^{2}
\end{equation}
\indent In equation~\ref{autocorrelation_function_secondary}, $a$, $\tau_{1}$, $\tau_{ww}$ and $\beta$ are the four fitting parameters corresponding to the relaxation strength, the fast relaxation time, the slow relaxation time and the stretching exponent respectively. $C(t)$ is a second order autocorrelation function, with the two terms within the brackets of equation~\ref{autocorrelation_function_secondary} giving the first order autocorrelation function $g^{(1)}(t)$. It is to be noted that in the study of most molecular glass formers, dielectric spectroscopy, which yields a first order correlation function due to dipole-dipole correlations, has been often used to study relaxation processes \cite{Debenedetti_Metastable_Liquids}.\\
\begin{figure*}[!t]
\begin{center}
\includegraphics[width=5.3in]{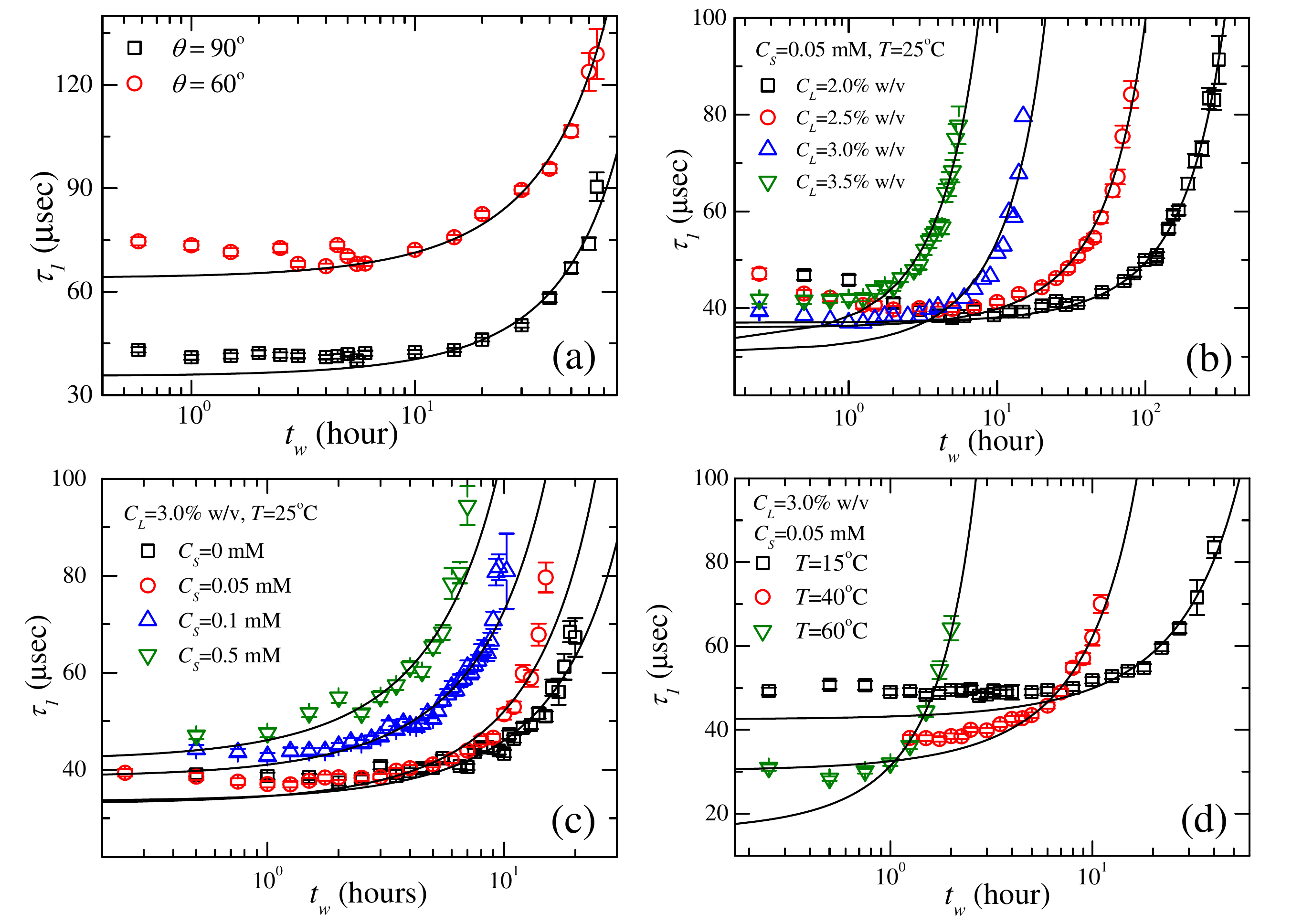}
\caption{(a) Fast relaxation time $\tau_{1}$ {\it vs.} waiting time $t_{w}$ for $C_{L}=$2.5\% w/v Laponite suspension ($C_{S}=0$ mM and $T=$25$^{o}$C) at two different scattering angles ($\theta=90^{o}$ ($\square$) and $\theta=60^{o}$ ($\circ$)). Solid lines are fits to equation~\ref{fast_relaxation_time_secondary}. $\tau_{1}$ is plotted {\it vs.} $t_{w}$ for Laponite suspensions with different (b) $C_{L}$ (2.0\% w/v ($\square$), 2.5\% w/v ($\circ$), 3.0\% w/v ($\triangle$) and 3.5\% w/v ($\nabla$)), (c) $C_{S}$ (0 mM ($\square$), 0.05 mM ($\circ$), 0.1 mM ($\triangle$) and 0.5 mM ($\nabla$)), and (d) $T$ (15$^{\circ}$C ($\square$), 40$^{\circ}$C ($\circ$) and 60$^{\circ}$C ($\nabla$)) values at scattering angle $\theta=90^{\circ}$.}
\label{fast_relaxation_secondary}
\end{center}
\end{figure*}
\indent We plot $\tau_{1}$ and $<\tau_{ww}>$ {\it vs.} $1/q^{2}$ in figures~\ref{autocorrelation_secondary}(b) and (c), where $q$ is the scattering wave vector at two different waiting times ($t_{w}=1$ hour and 30 hour) for a 2.5\% w/v Laponite suspension. A straight line passing through the origin indicates that both the relaxation processes at 25$^{\circ}$C are diffusive, i.e., $\tau_{1}=1/D_{1}q^{2}$ and $<\tau_{ww}>=1/D_{2}q^{2}$, where $D_{1}$ and $D_{2}$ are the diffusion coefficients corresponding to the fast and slow relaxation processes respectively. Another important observation is that both $\tau_{1}$ and $<\tau_{ww}>$ slow down with $t_{w}$.\\
\indent The aging dynamics of a spontaneously evolving Laponite suspension and its approach to a kinetically arrested state can be compared to molecular glasses which achieve the glass transition upon rapid cooling to avoid the crystalline state. It was demonstrated in an earlier work that Arrhenius and Vogel-Fulcher-Tammann (VFT) dependences of the fast and the slow relaxation times on waiting time respectively are obtained if the waiting time $t_{w}$ of a spontaneously evolving Laponite suspension is mapped with the inverse of the thermodynamic temperature $1/T$ of molecular glasses \cite{Debasish_YMJ_Ranjini_Soft_Matter_2014}. Consequently, the $t_{w}$-dependence of $\tau_{1}$ and $<\tau_{ww}>$ are given by the following equations \cite{Debasish_YMJ_Ranjini_Soft_Matter_2014}:
\begin{equation}
\tau_{1}=\tau^{0}_{1}\exp\left[\frac{t_{w}}{t^{\infty}_{\beta}}\right]
\label{fast_relaxation_time_secondary}
\end{equation}    
and,
\begin{equation}
<\tau_{ww}>=<\tau_{ww}>^{0}\exp\left[\frac{Dt_{w}}{t^{\infty}_{\alpha}-t_{w}}\right]
\label{slow_relaxation_time_secondary}
\end{equation}    
\indent In equation~\ref{fast_relaxation_time_secondary}, the fitting parameter $\tau^{0}_{1}$ is equal to $\tau_{1}(t_{w}\rightarrow 0)$ while $t^{\infty}_{\beta}$ is the characteristic timescale associated with the slowdown of the secondary relaxation process. In equation~\ref{slow_relaxation_time_secondary}, $D$, $<\tau_{ww}>^{0}$ and $t^{\infty}_{\alpha}$ are the three fitting parameters. $D$ is called the fragility or strength parameter which quantifies the deviation from an Arrhenius behavior, $<\tau_{ww}>^{0}=<\tau_{ww}>(t_{w}\rightarrow 0)$, and $t^{\infty}_{\alpha}$ is the Vogel time, or the waiting time at which $<\tau_{ww}>\rightarrow\infty$ \cite{Debasish_YMJ_Ranjini_Soft_Matter_2014}.\\
\indent In figure~\ref{fast_relaxation_secondary}(a), the time-evolutions of $\tau_{1}$ is plotted {\it vs.} $t_{w}$ for two scattering angles ($\theta=60^{\circ}$ and $90^{\circ}$) for a 2.5\% Laponite suspension. It can be seen from figure~\ref{fast_relaxation_secondary} that $\tau_{1}$ decreases with $t_{w}$ at smaller $t_{w}$, while at higher $t_{w}$, $\tau_{1}$ increases with $t_{w}$ for all the Laponite suspensions with different $C_{L}$, $C_{S}$ and $T$ investigated here. It is known that Laponite particles can form tactoids or rigid aggregates consisting of more than one platelet in aqueous suspension \cite{Samim_Ranjini_Langmuir_2013,Shahin_Joshi_Langmuir_2012}. At the early stage of dispersion, these tactoids exfoliate and the rate of exfoliation decreases with time as the intertactoid Coulombic repulsion increases rapidly with time \cite{Samim_Ranjini_Langmuir_2013}. This tactoid fragmentation process results in the speeding up of the dynamics and gives rise to the observed non-monotonicity in the $\tau_{1}$ {\it vs.} $t_{w}$ plot at small $t_{w}$ (figure~\ref{fast_relaxation_secondary}). It is seen from figure~\ref{fast_relaxation_secondary} that the later increasing part of $\tau_{1}$ can be fitted to equation~\ref{fast_relaxation_time_secondary} and the rate of increase becomes faster for higher values of $C_{L}$, $C_{S}$ and $T$.\\
\indent The diffusion coefficient $D_{1}$ of a sphere is related to its relaxation time $\tau_{1}$ by the relation $\tau_{1}=1/D_{1}q^{2}$ \cite{bern_pecora}. $D_{1}$ can be estimated for a dilute suspension of monodisperse spheres from the Stokes-Einstein relation, $D_{1}=k_{B}T/6\pi\eta r_{h}$ \cite{bern_pecora}, where $k_{B}$, $\eta$ and $r_{h}$ are the Boltzmann constant, viscosity of the medium (0.89 mPa.s at 25$^{\circ}$C) and hydrodynamic radius ($r_{h}=d_{s}/2$) of the particle respectively. It follows, therefore, that $\tau=1/D_{1}q^{2}=6\pi\eta r_{h}/k_{B}Tq^{2}$. As the Laponite platelet is a disk shaped particle (diameter $d=25-30$ nm and thickness $h=1$ nm approximately \cite{Kroon_PRE_1996}), its equivalent spherical diameter (ESD) is given by the Jennings-Parslow relation, $d_{s}=d\left(\frac{3\arctan\sqrt{(d/h)^{2}-1}}{2\sqrt{(d/h)^{2}-1}}\right)^{1/2}$ \cite{Jennings_Parslow_PRSLA_1988}. Using the Jennings-Parslow relation, the estimated value of $d_{s}$ for a Laponite platelet is $7.5-8.3$ nm. Given the value of $d_{s}=7.5-8.3$ nm for the Laponite platelet, the estimated value of the diffusion timescale of the platelet $\tau$ is $30-34$ $\mu$sec and $62-68$ $\mu$sec at $q=0.0223$ nm$^{-1}$ ($\theta=90^{\circ}$) and $q=0.0157$ nm$^{-1}$ ($\theta=60^{\circ}$) respectively. It is seen from figure~\ref{fast_relaxation_secondary}(a) that the values of $\tau^{0}_{1}$, i.e. $\tau_{1}(t_{w}\rightarrow 0)$, is 34 $\mu$sec and 63 $\mu$sec for $\theta=90^{\circ}$ and $\theta=60^{\circ}$ respectively, which are in excellent agreement with the estimated values of $\tau_{1}$ at smaller $t_{w}$. Therefore, $\tau_{1}$ can be associated with a relaxation process that involves a single Laponite platelet.\\
\begin{figure}
\begin{center}
\includegraphics[width=2.8in]{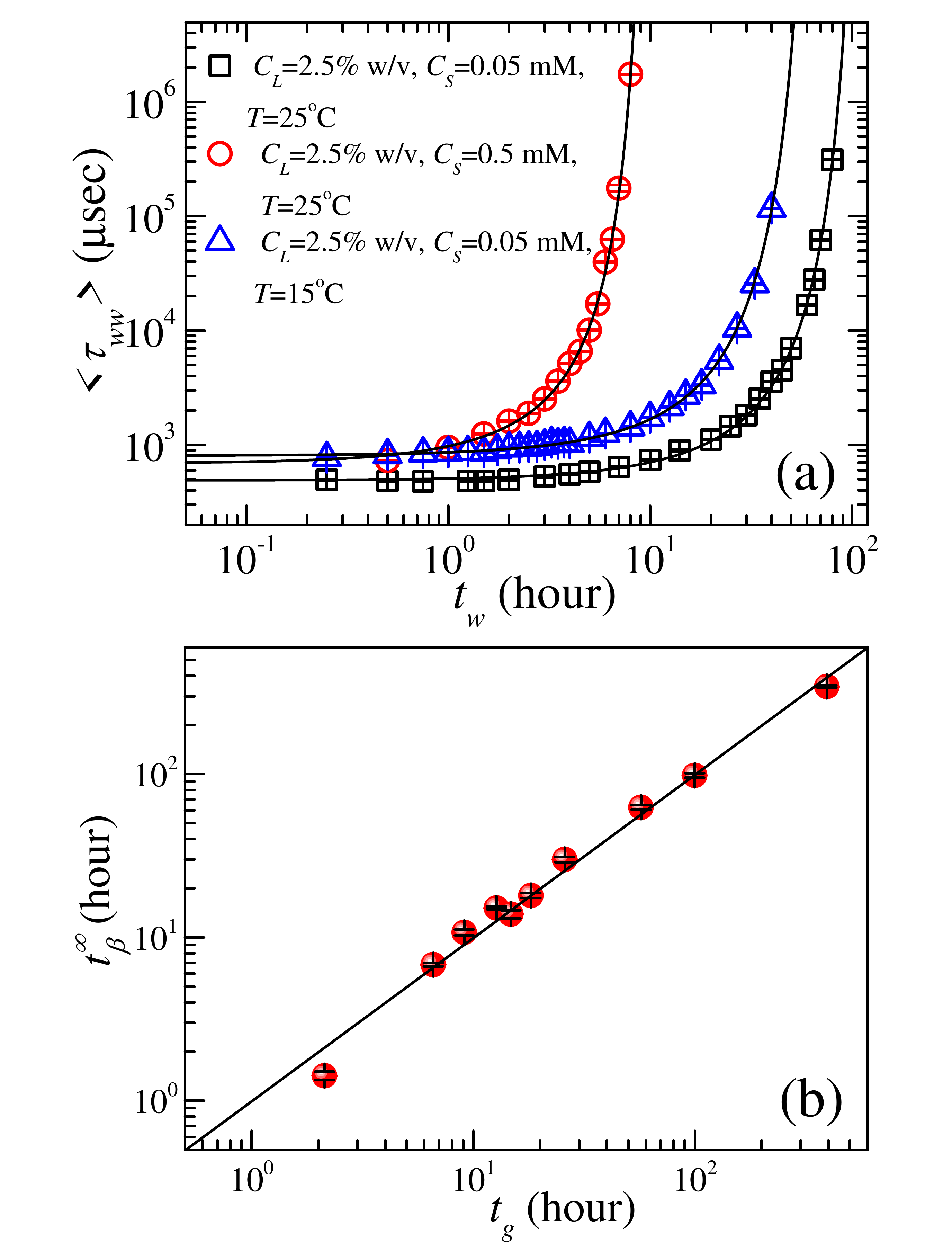}
\caption{(a) $<\tau_{ww}>$ {\it vs.} $t_{w}$ for Laponite suspensions with three different $C_{L}$, $C_{S}$ and $T$. Solid lines are fits to equation~\ref{slow_relaxation_time_secondary}. (b) The characteristic timescale associated with the secondary relaxation process $t^{\infty}_{\beta}$ {\it vs.} the ergodicity breaking time $t_{g}$ for different $C_{L}$, $C_{S}$ and $T$. The solid line ($t^{\infty}_{\beta}$ $\approx$ (0.99$\pm$0.05)$t_{g}$) is a linear fit passing through the origin.}
\label{glass_transition_time_secondary}
\end{center}
\end{figure}
%
\indent It was an empirical observation for many supercooled liquids that the glass transition temperature $T_{g}$ is proportional to the activation energy $E_{\beta}$ of the slow secondary $\beta$-relaxation process \cite{Kudlik_et_al_EPL_1997,Prevosto_et_al_Philosophical_Magazine_2008}. We define an ergodicity breaking time $t_{g}$ for Laponite suspensions from the time evolution of $<\tau_{ww}>$. In figure~\ref{glass_transition_time_secondary}(a), it can be seen that $<\tau_{ww}>$ increases with $t_{w}$ for all $C_{L}$, $C_{S}$ and $T$. It can also be seen that the time-evolution of $<\tau_{ww}>$ can be fitted to equation~\ref{slow_relaxation_time_secondary} very well. In supercooled liquids $T_{g}$ is defined as the temperature at which mean $\alpha$-relaxation time is 100 sec \cite{Angell_J_Non_Cryst_Solids_1991}. Similarly, for the Laponite suspensions studied here, $t_{g}$ is defined as the waiting time at which $<\tau_{ww}>=100$ sec and extrapolated from fits to equation~\ref{slow_relaxation_time_secondary}. In figure~\ref{glass_transition_time_secondary}(b), the characteristic timescale $t^{\infty}_{\beta}$ associated with the fast relaxation process and extracted from fits to equation~\ref{fast_relaxation_time_secondary} (figure~\ref{fast_relaxation_secondary}(b), (c) and (d)), is plotted {\it vs.} $t_{g}$ for Laponite suspensions with different $C_{L}$, $C_{S}$ and $T$ that were studied in this work. It is seen from the figure that $t^{\infty}_{\beta}\propto t_{g}$, indicating a possible coupling between the fast and slow relaxation processes. A linear fit to the data (shown by solid line) yields $t^{\infty}_{\beta}=(0.99\pm 0.05)t_{g}$. In the present context, the Arrhenius $t_{w}$-dependence of $\tau_{1}$ indicates a connection of $t^{\infty}_{\beta}$ with the activation energy $E_{\beta}$ of the secondary relaxation process, i.e. $t^{\infty}_{\beta}\propto 1/E_{\beta}$ \cite{Debasish_YMJ_Ranjini_Soft_Matter_2014}. Hence, the coupling between $t^{\infty}_{\beta}$ and $t_{g}$ ($t^{\infty}_{\beta}\propto t_{g}$) is reminiscent of the linear relationship between $E_{\beta}$ and $T_{g}$ in supercooled liquids \cite{Kudlik_et_al_EPL_1997,Prevosto_et_al_Philosophical_Magazine_2008}.\\
%
\begin{figure}[!t]
\begin{center}
\includegraphics[width=2.8in]{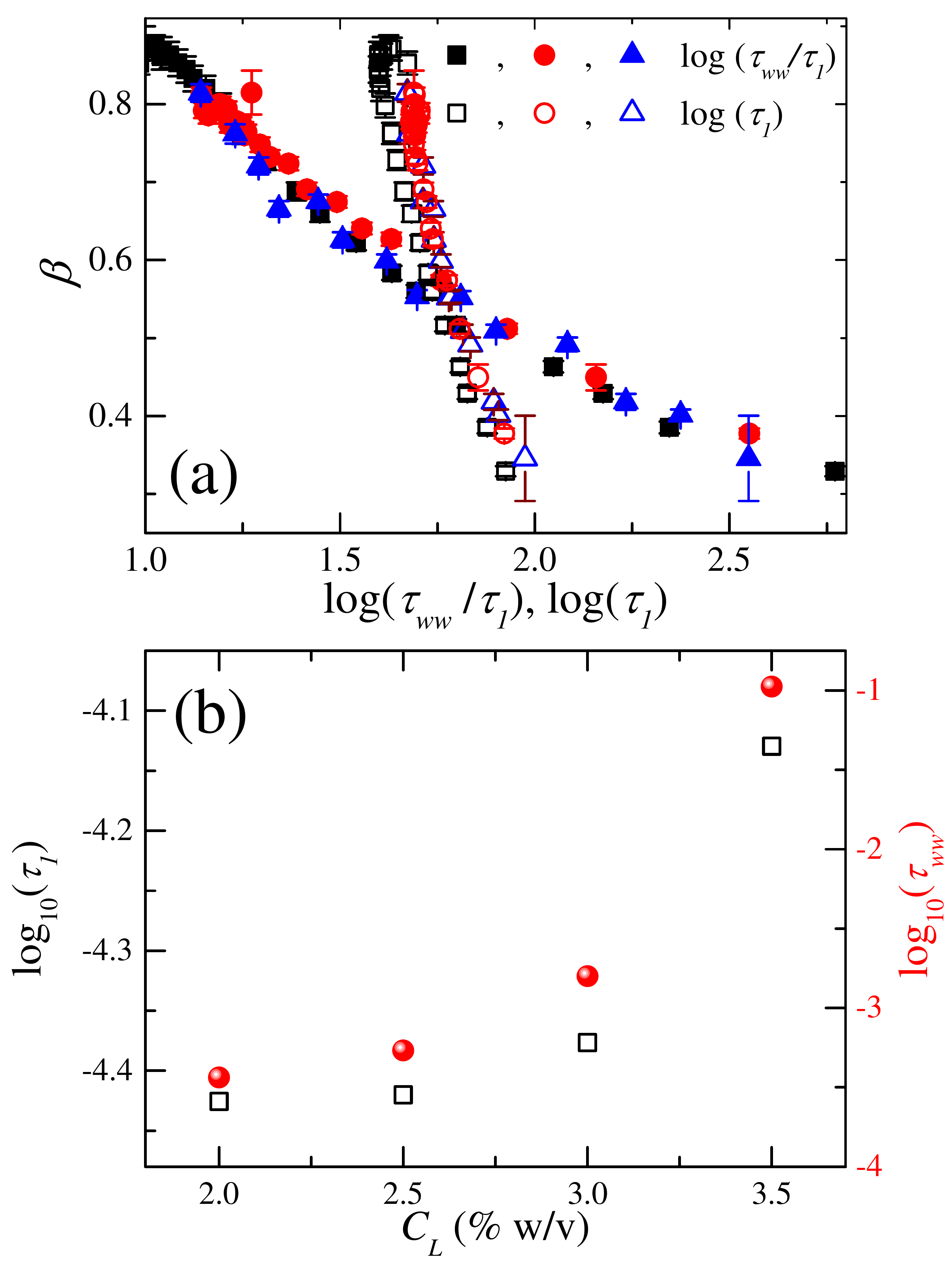}
\caption{(a) Stretching exponent $\beta$ {\it vs.} $\log{(\tau_{1})}$ (open symbols) and $\log{(\tau_{ww}/\tau_{1})}$ (solid symbols) for Laponite suspensions with three different $C_{L}$, $C_{S}$ and $T$ ($C_{L}=2.5\%$, $C_{S}=0.05$ mM and $T=25^{\circ}$C ($\square$, $\blacksquare$), $C_{L}=3.0\%$, $C_{S}=0.05$ mM and $T=15^{\circ}$C ($\circ$, $\bullet$) and $C_{L}=3.0\%$, $C_{S}=0.5$ mM and $T=25^{\circ}$C ($\triangle$, $\blacktriangle$)). (b) $\log{(\tau_{1})}$ ($\square$) and $\log{(\tau_{ww})}$ ($\bullet$) of Laponite suspensions of $t_{w}=5.5$ hour {\it vs.} $C_{L}$ ($C_{S}=0.05$ mM and $T=25^{\circ}$C).}
\label{width_secondary}
\end{center}
\end{figure}
\indent It was reported from another empirical observation that the width of the $\alpha$-relaxation process of a glass former is coupled with the J-G relaxation timescale, with $\beta$ decreasing monotonically with $\log{(\tau_{\beta})}$ or $\log{(\tau_{\alpha}/\tau_{\beta})}$ \cite{Fragiadakis_Roland_PRE_2013,Ngai_JCP_1998}. In the present context, the $\alpha$-relaxation process is a non-exponential process and is expressed in terms of a stretched exponential function (equation~\ref{autocorrelation_function_secondary}). The non-exponential decay arises due to a distribution of relaxation timescales that is given by the well-known Kohlrausch-Williams-Watts distribution: $\rho_{ww}(\tau)=-\frac{\tau_{ww}}{\pi\tau^2}\sum^{\infty}_{k=0}\allowbreak\frac{(-1)^k}{k!}\sin(\pi\beta k)\Gamma(\beta k+1)\left(\frac{\tau}{\tau_{ww}}\right)^{\beta k+1}$ \cite{Lindsey_JCP_1980}. The width $w$ of the distribution is related to the stretching exponent $\beta$ and is given by $w=\frac{<\tau^{2}_{ww}>}{<\tau_{ww}>^{2}}-1=\frac{\beta\Gamma(2/\beta)}{(\Gamma(1/\beta))^{2}}-1$. In figure~\ref{width_secondary}(a), we plot $\beta$ {\it vs.} $\log(\tau_{1})$ and $\log(\tau_{ww}/\tau_{1})$. It is seen that $\beta$ decreases with $\log(\tau_{1})$ and $\log(\tau_{ww}/\tau_{1})$ indicating that the width of the $\alpha$-relaxation process is indeed coupled with the secondary relaxation time \cite{Fragiadakis_Roland_PRE_2013}. It is also observed from figure~\ref{width_secondary}(a) that the superpositions of $\beta$ values for all $C_{L}$, $C_{S}$ and $T$ with $\log(\tau_{1})$ and $\log(\tau_{ww}/\tau_{1})$ is reminiscent of the self-similarity of the relaxation processes ($\tau_{1}$ and $<\tau_{ww}>$) for these physicochemical parameters ($C_{L}$, $C_{S}$ and $T$) reported earlier \cite{Debasish_YMJ_Ranjini_Langmuir_2015}.\\ 
\indent It was observed from dielectric measurements that the separation between the $\alpha$ and $\beta$ relaxation peaks is greater at elevated pressures for molecular glass formers, thus confirming the sensitivity of the $\beta$ relaxation process to high pressures \cite{Paluch_Roland_Pawlus_Ziolo_Ngai_PRL_2003}. In figure~\ref{width_secondary}(b), we plot $\log(\tau_{1})$ and $\log(\tau_{ww})$ for different $C_{L}$ for a fixed waiting time $t_{w}=5.5$ hour. It is seen from figure that both $\tau_{1}$ and $\tau_{ww}$ increase with $C_{L}$. Increasing $C_{L}$ is analogous to increasing pressure on the system as a larger number of particles are packed in a fixed volume. Clearly, both the relaxation processes of Laponite suspensions are sensitive to pressure or concentration.\\
\indent It is observed in a variety of glass formers that the order of magnitude of the J-G $\beta$-relaxation time $\tau_{\beta}$ lies very close to the primitive $\alpha$-relaxation time $\tau^{*}$ of the coupling model \cite{Ngai_Rendell_Rajagopal_Teitler_Ann_New_York_Acad_Sci_1986,Ngai_Rajagopal_Teitler_JCP_1988,Rajagopal_Teitler_Ngai_JPC_1984}. According to the coupling model, the relaxation process of a complex system depends on the interactions among the relaxation units. A relaxation unit relaxes independently with a rate $W_{0}=1/\tau_{0}$ at a shorter time $t$ than a crossover or characteristic time $t_{c}$. The correlation function $\phi(t)$ can be expressed by a simple exponential function, $\phi(t)=\exp{(-t/\tau_{0})}$. At longer time i.e. $t>t_{c}$, the primitive relaxation rate slows down and becomes time-dependent due to the many-body interactions or the coupling between relaxation units and can be expressed by $W(t)=W_{0}(t/t_{c})^{-n}=W_{0}(t/t_{c})^{\beta-1}$, where $n=1-\beta$ is the coupling parameter. It was also observed that $t_{c}$ is insensitive to the temperature of a molecular glass former. In the regime $t>t_{c}$, the correlation function can be described by a stretched exponential function $\phi(t)=\exp{[-(t/\tau^{*})^{\beta}]}$, where $\tau^{*}$ is the effective relaxation time and related to $\tau_{0}$ (or $W^{-1}_{0}$), the initial primitive relaxation time in the absence of coupling by the the following expression:\\
\begin{equation}
\tau^{*}=\left[\beta t^{\beta-1}_{c}W^{-1}_{0}\right]^{1/\beta}
\label{two_coupled_second}
\end{equation}
\indent The microscopic relaxation rate $W_{0}$ is related to the diffusion coefficient $D_{0}$ of a single relaxation unit and is given by $W_{0}=(D_{0}q^{2})$ with $D_{0}=d^{2}_{s}/\tau_{0}$. Therefore, equation~\ref{two_coupled_second} can be written in the following form \cite{Ngai_Coupling_Model_Macromolecules_1992}:
\begin{equation}
\tau^{*}=\left[\beta t^{\beta-1}_{c}\frac{\tau_{0}}{d^{2}_{s}q^{2}}\right]^{1/\beta}
\label{two_coupled_third}
\end{equation}
It was reported in an earlier work that $t_{c}$ is concentration dependent for Laponite suspensions and that the value of $t_{c}$ at higher concentrations ($C_{L}>1.5\%$ w/v) is approximately 20 $\mu$sec \cite{Zulian_et_al_J_Non_Cryst_Solids_2007}. Remarkably, the estimated value of the primitive relaxation time $\tau_{0}$ (i.e. $t<t_{c}$) from equation~\ref{two_coupled_third} is 10 $\mu$sec $<\tau_{0}<$ 20 $\mu$sec, which is of the same order of magnitude as $\tau_{1}$ (figure~\ref{fast_relaxation_secondary}). This result is reminiscent of the observation in molecular glass formers where the primitive relaxation time and J-G $\beta$-relaxation has almost the same order of magnitude. These observations indicate that the $\beta$-relaxation process of Laponite suspensions reported here could be a plausible candidate for the J-G $\beta$-relaxation process \cite{Ngai_Rendell_Rajagopal_Teitler_Ann_New_York_Acad_Sci_1986,Ngai_Rajagopal_Teitler_JCP_1988,Rajagopal_Teitler_Ngai_JPC_1984}. 
\section{Conclusions}
In this work, we have extracted the primary and secondary relaxation timescales of Laponite suspensions with different Laponite concentrations ($C_{L}$), salt concentrations ($C_{S}$) and temperatures ($T$) from intensity autocorrelation functions obtained from dynamic light scattering (DLS) measurements. A well-known route is followed to characterize the secondary relaxation that are extracted from our experiments \cite{Ngai_Paluch_JCP_2004}. It is observed that the secondary relaxation process of aging Laponite suspensions involves a single rigid Laponite particle. Furthermore, it is coupled with the primary relaxation process as it is observed that its ergodicity breaking time $t_{g}\propto t^{\infty}_{\beta}$, the characteristic timescale associated with the $\beta$-relaxation process. It is also demonstrated that the width $w$ of the distribution of the $\alpha$-relaxation timescale is correlated with the $\beta$-relaxation timescale. Both $\tau_{1}$ and $\tau_{ww}$ are found to be very sensitive to concentration within the range of $C_{L}$ explored here. Finally, it is observed that the order of magnitude of the primitive relaxation time $\tau_{0}$ estimated from the coupling model is same as that of the secondary $\beta$-relaxation time $\tau_{1}$. Our experimental observations therefore clearly suggest that the $\beta$-relaxation process of colloidal glasses of Laponite shows many characteristics of the J-G $\beta$-relaxation processes that are seen in molecular glass formers. Interestingly, the size of the Laponite particle (ESD $\approx 8$ nm) is only a little bigger than a molecule ($0.1$ nm $<$ ESD $<1$ nm). Laponite particles are also asymmetric in shape and have charge anisotropy. In general, these particles are categorized as patchy colloidal particles. Similarly, for a molecular glass former, the basic building block i.e. a molecule, in general, is asymmetric and possesses non-uniform charge distribution. Colloidal patchy particles of Laponite also form colloidal glasses at very low concentrations (volume fraction $<0.2$), which allows adequate volume for the whole particle to reorient and diffuse and take part in the relaxation processes. Therefore, all these factors (size, shape, concentration etc.) also indicate that the soft glassy suspensions of Laponite particles are excellent candidates among several well known colloidal glass formers to exhibit a J-G $\beta$-relaxation mode.       
\bibliography{research}
\end{document}